\documentstyle[12pt]{article}
\newcommand{\be}{\begin{equation}}
\newcommand{\ee}{\end{equation}}
\newcommand{\bea}{\begin{eqnarray}}
\newcommand{\eea}{\end{eqnarray}}
\newcommand{\beas}{\begin{eqnarray*}}
\newcommand{\eeas}{\end{eqnarray*}}
\newcommand{\bi}{\begin{itemize}}
\newcommand{\ei}{\end{itemize}}
\newcommand{\bc}{\begin{center}}
\newcommand{\ec}{\end{center}}
\newcommand{\bfl}{\begin{flushleft}}
\newcommand{\efl}{\end{flushleft}}
\newcommand{\bfr}{\begin{flushright}}
\newcommand{\efr}{\end{flushright}}
\newcommand{\f}{\frac}

\def\6{\partial} \def\a{\alpha} 
\def\g{\gamma} \def\d{\delta} 
\def\e{\epsilon}

\def\m{\mu}   \def\p{\pi}
 \def\s{\sigma} \def\t{\tau}
  
\def\o{\omega}  
 \def\L{\Lambda}

\begin{document}

\title{Unconstrained Variables of Non-Commutative Open Strings}

\author{M. A. De Andrade$^{a}$\footnote{E-mail: marco@cbpf.br, marco@gft.ucp.br}, 
M. A. Santos$^{b}$\footnote{masantos@gft.ucp.br} and I. V. Vancea$^{c}$
\footnote{E-mail: ivancea@unesp.ift.br, vancea@cbpf.br.On leave from Babes Bolyai 
University of Cluj, Romania.}}

\date{\small $^a$Grupo de F\'{\i}sica Te\'{o}rica, Universidade Cat\'{o}lica de Petr\'{o}polis \\ 
Rua Bar\~ao de Amazonas 124, 25685-070,  Petr\'opolis - RJ, Brasil \\ and \\
Departamento de F\'{\i}sica Te\'{o}rica, Universidade Federal de Juiz de Fora \\ 
Cidade Universit\'{a}ria, Caixa Postal 656, 36036-330, Juiz de Fora - MG, Brasil. \\
\vspace{.5cm}
$^b$Departamento de F\'{\i}sica, Universidade Federal Rural do Rio de Janeiro\\
23851-180, Serop\'{e}dica - RJ, Brasil\\ and \\
Centro Brasileiro de Pesquisas F\'{\i}sicas,\\
Rua Dr. Xavier Sigaud 150, 22290-180 Rio de Janeiro - RJ, Brasil\\
\vspace{.5cm} 
$^c$Instituto de F\'{\i}sica Te\'{o}rica , Universidade Estadual Paulista \\ 
Rua Pamplona 145, 015405-900.  S\~ao Paulo - SP, Brasil \\}

\maketitle

\abstract{The boundary conditions of the bosonic string theory in non-zero $B$-field 
background are equivalent to the second class constraints 
of a discretized version of the theory. By projecting the original canonical coordinates
onto the constraint surface we derive a set of coordinates of string that
are unconstrained. These coordinates represent a naturalframework for the quantization of
the theory.}

\newpage

\section{Introduction}

String theory provides new insights in the structure of fundamental interactions at
short distances that are unexpected from the point-particle physics. One of the most
intriguing aspects of string physics is the non-commutativity of space-time and of 
super Yang-Mills interactions which is a 
feature of backgrounds with $N$ parallel $D$-branes one atop of the other 
\cite{ew} or with bound states of branes and constant, non-vanishing  Kalb-Ramond 
field \cite{sw}. The non-commutativity of brane bound states is a consequence of
the non-commutativity of some of the coordinates of the open strings that end on
the branes. The effect of the $B$-field in this case is to mix the Neumann and
Dirichlet boundary conditions. The mixed boundary conditions constrain the fields, 
but only on the boundary. However, since the boundary conditions are necessary in 
order to give the bracket structure of the theory, one has to take them into account
in the quantization process. A similar situation arises in the case of constrained
systems, with the difference that there the constraints hold in the bulk, too.
For example, if the constraints are of second class
the canonical Poisson brackets are not compatible with them
and therefore must be replaced with the Dirac brackets which are compatible with the
constraints but are not canonical \cite{pamd}. This analogy was used in 
\cite{aas1,aas2,ss,ch1,ch2,tl,mz}
where the mixed boundary conditions were treated as primary and secondary Hamiltonian 
constraints. A non-conventional Dirac quantization was performed in \cite{aas2,ss}
where the canonical Hamiltonian was used instead of the primary Hamiltonian and 
non-commutative coordinates were shown to exist on the boundary as well as in the bulk.
In \cite{ch1} the same quantization procedure but with an infinite number of secondary
constraints led to non-commutative coordinates on the boundary only. 
A different approach was performed in \cite{ko} where a 
non-vanishing Lagrange multiplier for constraints was used. It was showed there that the
primary constraints form a second class constraint algebra which, upon quantization,
led to non-commutative coordinates on the boundary while the coordinates in the bulk
were shown to be commutative. These results show that the non-commutativity of at least
a part of the variables (bulk variables) depends on the quantization procedure. However, 
in \cite{ir} the BRST conversion procedure \cite{lf,fs} was employed to modify the second 
class constraints
to first class ones and to obtain an equivalent model which has only commuting target space 
coordinates. The point is that some of the string variables are subject to constraints 
which
represent the boundary conditions after discretizing the string. These constraints represent 
complicate 
relations among the canonical coordinates which are no longer independent.

The aim of this paper is to provide an unconstrained set of coordinates on the constraint surface. To 
do that, we are going to apply a projector (constructed with the elements of the Dirac matrix) to the original
coordinates. The result is a set of new coordinates that parametrize locally the constraint surface with 
a set of linear
algebraic relations among them which allow to define a set of independent variables which are true in the sense
that they encode completely the dynamics of the open string. The unconstrained variables do not require the
introduction of any bracket structure other than Poisson bracket and therefore they represent
a natural framework to perform the canonical quantization.
If we choose as dependent variables the set of coordinates that correspond to the two ends of the string, 
then the open string is equivalent
with a system for with some cyclic coordinates.

The outline of the paper is the following. In the Section 2 we review the basic ideas of the method we are going to
use and which was developed in \cite{pp,smp,mas,majh,mns}. (For a review we refer the reader to \cite{mmv2}.) In 
Section 3 we find the unconstrained coordinates of the open strings and the Hamiltonian that governs its dynamics.
The last section is devoted to discussions.

\section{Projected variables for second class constraints}

Consider a constrained system in the Hamiltonian formalism which has only second class constraints. We denote the
canonical conjugate variables by $Z^M = (X^\m , P_\m)$ and the constraints by $\o_i = \o_i (Z^M)$. The constraints
determine a surface in the phase space on which the dynamics of the system takes place. An important 
object of the theory is the matrix whose elements are the Poisson brackets of the constraints \cite{pamd}
\be
C_{ij} = \{\o_i\; ,\; \o_j \}
\label{cmatrix}
\ee 
which has a non-vanishing determinant on the constraint surface, and therefore it is invertible, although it may be 
singular outside it. We define its inverse by the usual relation 
\be
C_{ij}\,C^{j\,k}\;=\;\delta _i^k
\label{cinverse}.
\ee
In the phase space there are gradient vectors $n^i$ to the constraint surface whose components are 
$\delta \o_i / \delta Z^M$. Using the normal vectors one can construct a projector onto the surface
defined by the following relation \cite{pp,mmv2} 
\begin{equation}
\Lambda ^{MN}\; \equiv \;\delta
^{\,MN}-J^{ML}\,\frac{\delta \omega _{i}}{\delta Z^{L}}\,C^{ij}\,\frac{%
\delta \omega _{j}}{\delta Z^{N}}, 
\label{projdef}
\end{equation}
where $J^{MN}$ is the symplectic matrix in the phase space defined by the Poisson brackets among
the canonical variables
\begin{equation}
J^{MN}\;\equiv\;\{Z^M\;,\;Z^N\}
\label{simplmatrix}.
\end{equation}
Note that the projector $\L$ can also be written in terms of the Dirac matrix $D$ 
\begin{equation}
\Lambda ^{MN}\;=\;D^{MP}\,(J^T)^{PN},
\label{lambdaDJ}
\end{equation}
where the upper index $J^T$ is the transposed matrix.
The elements of $D$ are given by the Dirac brackets among the canonical variables
\begin{equation}
D^{MN}\; \equiv\; \{ Z^M\;,\;Z^N\}_D
\label{diracmatrix},
\end{equation}
where the Dirac brackets are defined by the following relation
\begin{equation}
\{A\;,\;B\}_{D}\;\equiv\; \{A\;,\;B\}-\{A\;,\;\omega _{i}\}\,C^{ij}\,\{\omega
_{j}\;,\;B\}.
\label{diracbrackg}
\end{equation}
By using the matrix $\L$ in either (\ref{projdef}) or (\ref{lambdaDJ}) form, one can project
the canonical variables onto the constraint surface
\be
Z^* = \L Z
\label{projectg}.
\ee
The projected variables $Z^*$ are still subject to
some mutual relations which usually take a simpler form than the original constraints. When these
relations are linear, the dynamics is determined only by some local variables  
on the constraint surface. These variables are linearly independent. The dependent ones
can be eliminated by solving the simple algebraic relations \cite{pamd} locally. 
The projector (\ref{projdef}) helps us to find these "true" variables. However, we should note
that in practice it is not possible to apply this method always. The major obstacle is that
the Dirac brackets may be difficult to compute. In what follows we will find the unconstrained
variables of the open bosonic string in a two-form background to which this method applies
straightforwardly.

\section{Projected variables of the Non-Commutative \\ String}

Let us consider a bosonic open string in a closed string background with a non-vanishing $U(1)$ gauge potential 
$A^i(X)$ with a constant field-strength $F_{ij}$, a constant metric $g_{ij }$ and a
constant Kalb-Ramond field $B_{ij}(X)$. The action of the system is given by the following relation
\bea
S= \frac{1}{4\pi \alpha ^{\prime }}\int d^{2}\sigma \left[ \partial
_{a}X^{i}\partial ^{a}X_{i}+2\pi \alpha ^{\prime }B_{ij}\varepsilon
^{ab}\partial _{a}X^{i}\partial _{b}X^{j}\right]
\nonumber\\ 
+\int d\sigma \d(\s - \p)A_{i}\partial
_{\sigma }X^{i}
-\int d\sigma \d(\s )A_{i}\partial _{\sigma
}X^{i},  
\label{action}
\eea
where $\s^{\a} = (\t , \s )$ are the world-sheet coordinates and $\e^{ab}$ is the anti-symmetric symbol in
two dimensions $a,b=0,1$. We choose to work in the $d$-dimensional Euclidean space-time for convenience and the 
target-space indices are $i,j=1,2, \ldots ,d$, where $d$ is even and for critical string $d=26$. Since the
gauge field is constant, one can conveniently set it to zero $F_{ij}=0$. By varying the action (\ref{action})
with respect to the fields one obtains the equations of motion
\begin{equation}
\partial _{a}\partial ^{a}X^{i}=0,  
\label{eqmotion}
\end{equation}
which hold if and only if the following mixed Dirichlet and Neumann boundary conditions hold
\be
g_{ij}\partial _{\sigma }X^{j}+2\pi \alpha ^{\prime }B_{ij}\partial _{\sigma
}X^{j}|_{\sigma =0,\pi} =0 
\label{boundcond}.
\ee 
The boundary conditions (\ref{boundcond}) represent relations among the momenta of string. In the usual
quantization process, these conditions should be imposed on the Hilbert space of the theory. However,
there is an alternative approach to quantization of this system, in which the boundary conditions are
treated as algebraic constraints \cite{aas1,aas2,ss,ch1}. To this end, one has to discretize firstly the
string by dividing the range of $\s$ parameter by $\e>0$. The coordinate $X^i(\s)$ for $i=1,2,\ldots, d$ 
is equivalent to the following discrete set $X^{i}_{\a}$ where $\a = 1,2, \ldots, m$ and $\e = \p /m$.
It is easy to see that the discretized Lagrangian has the following form
\be
L=\left( 4\pi \alpha ^{\prime }\right) ^{-1}\sum\limits_{\alpha }\left[
\varepsilon \left( \stackrel{.}{X}_{\alpha }^{i}\right) ^{2}-\frac{1}{%
\varepsilon }\left( X_{\alpha +1}^{i}-X_{\alpha }^{i}\right) ^{2}+4\pi
\alpha ^{\prime }B_{ij}\stackrel{.}{X}_{\alpha }^{i}\left( X_{\alpha
+1}^{j}-X_{\alpha }^{j}\right) \right]
\label{lagrangian},
\ee
while the boundary conditions (\ref{boundcond}) at $\s =0$ become
\be
\frac{g_{ij}}{\varepsilon }\left( X_{2}^{j}-X_{1}^{j}\right) +2\pi \alpha
^{\prime }B_{ij}\stackrel{.}{X}_{1}^{j}=0 
\label{discretbc}.
\ee
Similarly, an identical set of discretized  boundary conditions is obtained at $\s= \p$ with 
$ \{ 1,2 \}$ in (\ref{boundcond}) replaced by $\{ m,m-1 \}$.

Due to the discretization, the boundary conditions are equivalent to algebraic constraints on the variables
of the ends of the string and on their first neighbors, while the bulk variables are unconstrained
\cite{aas1,ko}. If we pass to the canonical conjugate variables $X^{i}_{\a}$ and $P_{i\a}$, we can easily 
check out that the constraints which take the following form 
\be
w_{i}=\frac{1}{\varepsilon }\left[ \left( 2\pi \alpha ^{\prime }\right)
^{2}B_{ij}P_{1}^{j}-\left( 2\pi \alpha ^{\prime }\right)
^{2}B_{ij}B^{jk}\left( X_{2k}-X_{1k}\right) +g_{ij}\left(
X_{2}^{j}-X_{1}^{j}\right) \right] \approx 0 . 
\label{constraints}
\ee
are of second class. There is a second set of constraints $\g_i$ at the other end of the string, which 
can be obtained from (\ref{constraints}) by replacing the corresponding coordinates and momenta. 
In the subsequent analysis, we can eliminate the parameter $\e$.

In order to construct the projector onto the constraint surface we organize the variable matrices
$X^{i}_{\a}$ and $P_{i\a}$ into a column vector by $Z^{M}$. It is easy to see that the correspondence between 
the indices $\{i,\a \} $ and $\{ M \}$ is given by the following relation
\be
M =\left( \mu +\alpha \right) +\left( m-1\right) \left( \mu -1\right) -1,
\label{indexrelation}
\ee
where $\m = i, d + i$ and $i = 1, 2, \ldots, d$. It is also useful to keep track of the canonical coordinates 
and their conjugate momenta variables in this notation. From the formula (\ref{indexrelation}) one can
see that the correspondence between the two sets of variables is  
\begin{equation}
X_{\alpha }^{i},P_{\alpha i}\rightarrow Z^{m\left( i-1\right) +\alpha
},Z_{m\left( d+i-1\right) +\alpha } 
\label{notation}. 
\ee
Next, we split the canonical variables in boundary variables and bulk variables, respectively, by
splitting the index $\a$ as follows $\a = 1, 2, n, m-1, m$, where $n=3,4, \ldots, m-2$.  It is
straightforward to compute the Dirac brackets for the discretized system (see, for example, \cite{ko}.) 
The inverse of the constraint matrix is given by the following relation
\begin{equation}
\left( C^{-1}\right) ^{ij}=\frac{\varepsilon ^{2}}{2\left( 2\pi \alpha
^{\prime }\right) ^{2}}\left[ \frac{1}{\left( g+2\pi \alpha ^{\prime
}B\right) }\frac{1}{B}\frac{1}{\left( g-2\pi \alpha ^{\prime }B\right) }%
\right] ^{ij}.
\label{inverseC}
\end{equation} 
From this, we can write down the $\L$-matrix by using
the formula (\ref{lambdaDJ}) with  the variables (\ref{notation}).
In order to write its non-vanishing components
we see from (\ref{indexrelation}) that the index $M$ can be denoted
by $i,\a$ and $d+i,\a$. Also, it is important to observe that the projector matrix
consists in four blocks according to the type of $Z$'s coordinates on which they act and which
they map into, i. e. $X$-like and $P$-like coordinates, respectively.
The non-vanishing components from the first block of  
$\L$-matrix are the following ones
\bea
2\L^{i(1)}_{\;j(1)} &=& 2 \L^{i(1)}_{\;j(2)} = 
\L^{i(2)}_{\;j(2)} = \L^{i(m-1)}_{\;j(m-1)} = 2\L^{i(m)}_{\;j(m-1)} = 
2\L^{i(m)}_{\;j(m)} = \d^{i}_{j}.\nonumber\\
\L^{i(n)}_{\;j(n')} &=& \d_{nn'}\d^{i}_{j}
\label{Lmatrixone}
\eea
Note that the index $\a$ is not a covariant index, therefore its relative position is irrelevant. It just
labels the different bulk and boundary discrete coordinates of string. The block (\ref{Lmatrixone}) maps
the position coordinates into position coordinates. In the same way we can write the non zero components
from the other blocks
\be
\L^{i(1)~d+j(1)} = - \L^{i(m)~d+j(m)} =
\f{1}{2}(2\pi \a')^2 
\left( \frac{1}{ \left( g+2\pi \alpha' B \right)} B \frac{1}{\left( g-2\pi \alpha'B \right)} \right)^{ij}
\label{Lmatrixtwo},
\ee
for the block that maps $P$'s into $X$'s,
\bea
\L_{d+i(1)~j(1)} &=& - \L_{d+i(1)~j(2)} = \L_{d+i(2)~j(1)} = \L_{d+i(2)~j(2)}  
= - \L_{d+i(m-1)~j(m-1)}\nonumber\\
&=& - \L_{d+i(m-1)~j(m)} = \L_{d+i(m)~j(m-1)} = - \L_{d+i(m)~j(m)}
\nonumber\\
&=& \f{1}{2(2\pi \a')^2} 
\left( \left( g+2\pi \alpha' B \right)\frac{1}{B}\left( g-2\pi \alpha'B \right) \right)_{ij}
\label{Lmatrixthree},
\eea
for the block that maps $X$'s into $P$'s and
\bea
2\L_{d+i(1)}^{\;\;\;d+j(1)} &=& 2 \L_{d+i(1)}^{\;\;\;d+j(2)} = 
\L_{d+i(2)}^{\;\;\;d+j(2)} = 
\L_{d+i(m-1)}^{\;\;\;d+j(m-1)} = 2 \L_{d+i(m)}^{\;\;\;d+j(m-1)} =
2 \L_{d+i(m)}^{\;\;\;d+j(m)} = \d^{i}_{j},\nonumber\\
\L_{d+i(n)}^{\;\;\;d+j(n')} &=& \d_{nn'}\d^{i}_{j}
\label{Lmatrixfour}
\eea
for the block that maps $P$'s into $P$'s. The elements of the projector are matrices with elements indexed
by $i,j$. We have chosen to work with the canonical structure of indices, i.e. $(X^i, P_j)$. Then it is
important that the $\L$-matrix maintains this structure through projection. Otherwise, we should include
in it the metric $g_{ij}$ in order to raise and lower the indices. This will change slightly
the form of the Dirac matrix. These complications are avoided by the above construction and the $\L$-matrix
given in (\ref{Lmatrixone})-(\ref{Lmatrixfour}) is consistent with the covariance of the phase space . 

By acting  with the $\L$-matrix on the coordinates of the phase space we arrive at the projected
variables on to the constraint surface. From (\ref{Lmatrixone})-(\ref{Lmatrixfour}) we obtain the following
projected coordinates
\begin{eqnarray}
Z^{\star m\left( i-1\right) +1} &=&\frac{1}{2}Z^{m\left( i-1\right) +1}+
\frac{1}{2}Z^{m\left( i-1\right) +2}
\nonumber\\
&+&\frac{1}{2}T^{-2}\left(
C^{-1}BA^{-1}\right) ^{ij}Z_{m\left( d+j-1\right) +1}  \label{13-1} 
\label{Z1}\\
Z^{\star m\left( i-1\right) +2} &=&Z^{m\left( i-1\right) +2}  \label{13-2}
\label{Z2} \\
Z^{\star m\left( i-1\right) +n} &=&Z^{m\left( i-1\right) +n}  \label{13-3}
\label{Z3} \\
Z^{\star m\left( i-1\right) +\left( m-1\right) } &=&Z^{m\left( i-1\right)
+\left( m-1\right) }  \label{13-4}
\label{Z4} \\
Z^{\star m\left( i-1\right) +m} &=&\frac{1}{2}Z^{m\left( i-1\right) +m}+
\frac{1}{2}Z^{m\left( i-1\right) +\left( m-1\right) }\nonumber\\
&-&\frac{1}{2}
T^{-2}\left( C^{-1}BA^{-1}\right) ^{ij}Z_{m\left( d+j-1\right) +m}
\label{Z5}\\
Z_{m\left( d+i-1\right) +1}^{\star } &=&\frac{1}{2}Z_{m\left( d+i-1\right)
+1}-\frac{1}{2}T^{2}\left( CB^{-1}A\right) _{ij}Z^{m\left( j-1\right) +2}
\nonumber\\
&+&
\frac{1}{2}T^{2}\left( CB^{-1}A\right) _{ij}Z^{m\left( j-1\right) +1}
\label{Z6} \\
Z_{m\left( d+i-1\right) +2}^{\star } &=&Z_{m\left( d+i-1\right) +2}+\frac{1}{
2}T^{2}\left( CB^{-1}A\right) _{ij}Z^{m\left( j-1\right) +2}
\nonumber\\
&+&\frac{1}{2}
T^{2}\left( CB^{-1}A\right) _{ij}Z^{m\left( j-1\right) +1}
+\frac{1}{2} Z_{m\left( d+i-1\right) +1}
\label{Z7} \\
Z_{m\left( d+i-1\right) +n}^{\star } &=&Z_{m\left( d+i-1\right) +n}
\label{Z8} \\
Z_{m\left( d+i-1\right) +m-1}^{\star } &=&Z_{m\left( d+i-1\right) +m-1}-
\frac{1}{2}T^{2}\left( CB^{-1}A\right) _{ij}Z^{m\left( j-1\right) +m-1}
\nonumber\\
&+&
\frac{1}{2}T^{2}\left( CB^{-1}A\right) _{ij}Z^{m\left( j-1\right) +m}
+\frac{1}{2}Z_{m\left( d+i-1\right) +m}
\label{Z9} \\
Z_{m\left( d+i-1\right) +m}^{\star } &=&-\frac{1}{2}Z_{m\left( d+i-1\right)
+m}
-\frac{1}{2}T^{2}\left( CB^{-1}A\right) _{ij}Z^{m\left( j-1\right) +m}
\nonumber\\
&+&\frac{1}{2}T^{2}\left( CB^{-1}A\right) _{ij}Z^{m\left( j-1\right) +m-1}
\label{Z10},
\end{eqnarray}
where we have used the following shorthand notations
\bea
A_{ij} &=& ( g-2\pi \alpha'B ) _{ij}
\nonumber\\
B_{ij} &=& B_{ij}
\nonumber\\
C_{ij} &=& ( g+2\pi \alpha' B ) _{ij}
\label{notationABC}
\eea
and $T=(2\pi\a')^{-1}$. Note that not all of the projected variables from (\ref{Z1})-(\ref{Z10}) are
independent. However, the relations among them are linear. We can show that there are two dependent
variables on the constraint surface. By using the symmetry with respect to the exchange of the ends of
the string one can easily find out these relation. We can pick up the coordinates on the boundary as
dependent variables and we obtain the following relations for them in terms of the linear independent 
coordinates
\begin{eqnarray}
Z^{\star m\left( i-1\right) +1} &=&T^{-2}\left( C^{-1}BA^{-1}\right)
^{ij}Z_{m\left( d+i-1\right) +1}^{\star }+Z^{\star m\left( i-1\right) +2}
\label{relone} \\
Z^{\star m\left( i-1\right) +m} &=&-T^{-2}\left( C^{-1}BA^{-1}\right)
^{ij}Z_{m\left( d+i-1\right) +m}^{\star }+Z^{\star m\left( i-1\right) +m-1}
\label{reltwo}.
\end{eqnarray}
From the relations above, we see that the system is equivalent with a system with $2d(m-1)$ independent
variables that parametrize locally the constraint surface. These variables commute among each other in agreement
with the result of \cite{ir}, but in this description the system is cyclic in the boundary coordinates which
implies that the corresponding momenta are conserved. The Hamiltonian of has the following expression
\begin{eqnarray}
H^{\star } &=&\frac{1}{4\pi \alpha ^{\prime }\varepsilon }
\sum\limits_{n=3}^{m-2}
\left[ \left( 2\pi \alpha ^{\prime }\right)^{2}
\left( Z_{m\left( d+i-1\right) +n}^{\star }-B_{ij}\left( Z^{\star
m\left( j-1\right) +n+1}
-Z^{\star m\left( j-1\right) +n}\right) \right)
^{2} \right.
\nonumber\\
&+&\left. \left( Z^{\star m\left( j-1\right) +n+1}-Z^{\star m\left( j-1\right)
+n}\right) ^{2}\right] 
\nonumber \\
&+&\frac{4(\pi\alpha')^{3} }{ \varepsilon }
\left[\left( C^{-1}BA^{-1}\right) ^{ij}Z^{\star}_{m \left( d+j-1 \right) +m }\right]^{2} 
+\frac{ 4(\pi\a')^3}{\varepsilon }\left[ \left( C^{-1}BA^{-1}\right) ^{ij}Z_{m\left(
d+j-1\right) +1}^{\star }\right] ^{2}  
\nonumber\\
&+& \frac{\pi \alpha'}{\varepsilon }
\left[ Z_{m\left( d+i-1\right) +1}^{\star } + 
\frac{1}{\left( 2\pi \alpha ^{\prime }\right) ^{-2}}
B_{ij} 
\left( C^{-1}BA^{-1}\right)
^{ij} Z_{m\left( d+j-1\right) +1}^{\star }
\right] ^{2} 
\nonumber \\
&+&\frac{\pi\a'}{\varepsilon }
\left[ Z_{m\left( d+i-1\right) +m-1}^{\star } + B_{ij}\left( 
\left( 2\pi \alpha ^{\prime }\right) ^{2}
\left(
C^{-1}BA^{-1}\right) ^{ij}Z_{m\left( d+j-1\right) +m}^{\star }\right)
\right] ^{2}.  
\label{hamiltonian}
\end{eqnarray}
We note that the coordinates of the ends of the string do not appear in (\ref{hamiltonian}). 
However, the corresponding momenta appear in the last four terms. The Hamiltonian (\ref{hamiltonian})
is written in terms of independent variables. Therefore, it represents the natural
functional for quantization.   

\section{Conclusions}

In this paper we have constructed the independent variables of the bosonic open string 
interacting with a constant $B$-field. The projected variables are described by the relations
(\ref{Z1})-(\ref{Z10}) and they parametrize locally the constraint surface. Through the
projection the constraints are reduced to linear relations among coordinates,
which is natural once the system is mapped onto the constraint surface. These relations
can be solved algebrically as we did in (\ref{relone}) and (\ref{reltwo}) and we are left
with the set of linearly independent variables. The dependent variables we have chosen
are the coordinates of the string on the boundary. In this parametrization the system is 
equivalent to a commutative string in the bulk which is cyclic on the coordinates on the
boundary. We note that after solving the relation (\ref{relone}) and (\ref{reltwo}) we
are left with the correct number of degrees of freedom corresponding to $2dm$ total variables
and $2d$ original constraints. On the constraint surface,
the dynamics of the system is described by the Hamiltonian given in (\ref{hamiltonian}).
Since we have a free theory which is cyclic on some variables on the constraint surface,
it is natural to use the canonical quantization in these parametrization in order to find
its spectrum.  

{\bf Acknowledgments}

M. A. S. and I. V. V. would like to thank to J. A. Helayel-Neto for hospitality at GFT-UCP during
the preparation of this work. I. V. V. would like to acknowledge to N. Berkovits for useful 
discussion and to FAPESP for partial support within a postdoc fellowship.


\begin{thebibliography}{99}

\bibitem{ew} E. Witten, Nucl. Phys. B460(1996)335, hep-th/9510135

\bibitem{sw} N. Seiberg and E. Witten, JHEP 09(1999)032, hep-th/9908142

\bibitem{pamd} P. A. M. Dirac, {\em Lecture Notes on Quantum Mechanics}, Yeshiva University Press
New York, 1964

\bibitem{aas1} F. Ardalan, H. Arfaei and M. M. Sheikh-Jabbari, JHEP 9902(1999)016

\bibitem{aas2} F. Ardalan, H. Arfaei and M. M. Sheikh-Jabbari, hep-th/9906161

\bibitem{ss} M. M. Sheikh-Jabbari and A. Shirzad, hep-th/9907055

\bibitem{ch1} C.-S. Chu and P.-M. Ho, Nucl. Phys. B550(1999)151

\bibitem{ch2} C.-S. Chu and P.-M. Ho, Nucl. Phys. B568(2000)447

\bibitem{tl} T. Lee, Phys. Rev. D62(2000)024022

\bibitem{mz} M. Zabzine, hep-th/0005142

\bibitem{ko} W. T. Kim and J. J. Oh, Mod. Phys. Lett. A15(2000)1597

\bibitem{ir} I. Rudychev, hep-th/0101039

\bibitem{lf} L. Fadeev, Phys. Lett. B145(1984)81

\bibitem{fs} L. Fadeev, S. Shatashvili, Phys. Lett. B167(1986)225 

\bibitem{pp} C. Marcio do Amaral, Nuovo Cim. B25(1975)817

\bibitem{smp} M. A. Santos, J. C.Mello and P. Pitanga Z. Phys. C55(1992)271

\bibitem{mas} M. A. Santos, J. C.Mello and P. Pitanga, Braz. J. Phys. 23(1993)214

\bibitem{majh} M. A. Santos and J. A. Helay\"{e}l-Neto, hep-th/9905065

\bibitem{mns} L. R. U. Manssur, A. L. M. A. Nogueira and M. A. Santos, hep-th/0005214

\bibitem{mmv2} M. A. De Andrade, M. A. Santos and I. V. Vancea, {\em Symplectic Projectors for Constrained Systems},
in preparation

\end{thebibliography}
\end{document}